\documentclass[lettersize,journal]{IEEEtran}

\usepackage{amsmath, amssymb, amsfonts}
\usepackage{algorithmic}
\usepackage{algorithm}
\usepackage{array}
\usepackage{graphicx}
\usepackage{cite}
\usepackage{textcomp}
\usepackage{url}
\usepackage{xcolor}
\usepackage{verbatim}
\usepackage{stfloats}
\usepackage{subfig}
\usepackage{caption}
\usepackage{makecell}
\usepackage{booktabs}
\usepackage{tabularx}
\usepackage{adjustbox}
\usepackage{hyperref}
\usepackage{multirow}
\usepackage{multicol}
\usepackage{makecell}

\begin{document}

\title{HuLA: Prosody-Aware Anti-Spoofing with Multi-Task Learning for Expressive and Emotional Synthetic Speech}

%\title{Is Expressive Synthetic Speech a Weakness for Anti-Spoofing? \\ Prosody-Aware Spoof Detection with HuLA}

\author{Aurosweta Mahapatra~\IEEEmembership{Student Member,~IEEE},  Ismail Rasim Ulgen~\IEEEmembership{Student Member,~IEEE,}  \\Berrak~Sisman~\IEEEmembership{Senior Member,~IEEE}
        % <-this % stops a space
\thanks{A. Mahapatra is with the Department of Electrical and Computer Engineering, Johns Hopkins University, Baltimore, MD 21218 USA (e-mail: amahapa2@jhu.edu).}
\thanks{Ismail Rasim Ulgen is with the Department of Electrical and Computer Engineering, Johns Hopkins University, Baltimore, MD 21218 USA (e-mail: iulgen1@jhu.edu).}
\thanks{B. Sisman is with the Department of Electrical and Computer Engineering, Johns Hopkins University, Baltimore, MD 21218 USA (e-mail: sisman@jhu.edu).}
\thanks{Manuscript received September 24, 2025.}}

\maketitle

\begin{abstract}
Current anti-spoofing systems remain vulnerable to expressive and emotional synthetic speech, since they rarely leverage prosody as a discriminative cue. Prosody is central to human expressiveness and emotion, and humans instinctively use prosodic cues such as $F_0$ patterns and voiced/unvoiced structure to distinguish natural from synthetic speech. In this paper, we propose \textit{HuLA}, a two-stage prosody-aware multi-task learning framework for spoof detection. In Stage~1, a self-supervised learning (SSL) backbone is trained on real speech with auxiliary tasks of $F_0$ prediction and voiced/unvoiced classification, enhancing its ability to capture natural prosodic variation similar to human perceptual learning. In Stage~2, the model is jointly optimized for spoof detection and prosody tasks on both real and synthetic data, leveraging prosodic awareness to detect mismatches between natural and expressive synthetic speech. Experiments show that HuLA consistently outperforms strong baselines on challenging out-of-domain dataset, including expressive, emotional, and cross-lingual attacks. These results demonstrate that explicit prosodic supervision, combined with SSL embeddings, substantially improves robustness against advanced synthetic speech attacks.

\end{abstract}
\begin{IEEEkeywords}
Anti-Spoofing, Multi-task Learning, Prosody.
\end{IEEEkeywords}

\section{Introduction}
Anti-spoofing aims to detect audio generated through replay attacks, speech synthesis, and voice conversion (VC) \cite{AS_Survey}. Recent progress in text-to-speech (TTS) \cite{Emoqtts, prompttts++, F5TTS, StyleTTS2, Emo-ctrlTTS} and VC systems \cite{expressive_vc, expressive_vc2, expressive_vc3, ESD, zkun1} has amplified concerns about expressive synthetic speech, which can be misused to compromise biometric authentication or impersonate speakers for spreading misinformation \cite{EthicalChallenge,misinfosurvey}. These risks highlight the urgent need for robust and generalizable anti-spoofing systems.

One of the goals of speech generation is to produce speech that is natural and indistinguishable from human speech. Expressiveness and emotion are the defining characteristics of human speech. TTS and VC systems often rely on prosody, particularly the fundamental frequency ($F_0$) to approximate these aspects \cite{F0_vc,F0_vc2,F0-synthesis,F0-synthesis2}. However, accurately modeling $F_0$ remains challenging \cite{chanllengeP-TTS, challengeP-VC2, challengeP-VC}, and current synthesis systems do not fully capture the subtleties of human expressiveness. While this limitation is a weakness for synthesis, it represents a valuable opportunity for anti-spoofing: imperfect expressiveness can serve as a discriminative cue for detecting synthetic speech.

Anti-spoofing research has been driven largely by community challenges such as ASVspoof \cite{ASVspoof2015} and the Audio Deep Synthesis Detection (ADD) Challenge \cite{ADD2022, ADD2023}, which introduced benchmark datasets including ASVspoof 2019 \cite{ASVspoof2019}, 2021 \cite{ASVspoof2021}, and 2024 \cite{ASVspoof2024}. These resources have significantly advanced the field and now serve as standard evaluation benchmarks.  

Current approaches span three main categories: (i) end-to-end models that operate directly on raw waveforms (e.g., RawNet2 \cite{RawNet2}, AASIST \cite{AASIST}); (ii) methods based on handcrafted spectral features combined with supervised classifiers (e.g., LCNN \cite{LCNN}, ASSERT \cite{ASSERT}, ResNet34 \cite{ResNet34}); and (iii) pretrained self-supervised learning (SSL) models such as wav2vec 2.0, HuBERT, and WavLM \cite{xlsr, wav2vec2, wavlm, hubert}. SSL-based approaches are now widely preferred due to their pretraining on large-scale data, ability to learn generalizable representations, and flexibility for fine-tuning in anti-spoofing tasks \cite{ssl_sls, ssl_rawnet2, xlsr-mamba}. Despite these advances, current SSL-based models fail to address the imperfect replication of emotion and expressiveness in synthetic speech, an underexplored, discriminative cue. In this work, we focus on the prosodic dimension of expressiveness and explicitly incorporate prosodic information alongside SSL embeddings to enhance spoof detection.

Prosody refers to suprasegmental features of speech \cite{prosody, prososdy2} that shape perception beyond linguistic content, playing a central role in conveying expressiveness and emotion. Humans are inherently sensitive to prosodic cues due to lifelong exposure to natural emotional and expressive speech, and can often distinguish real from synthetic audio based on these cues \cite{humanperception-DF}. Common prosodic attributes include fundamental frequency ($F_0$), jitter, shimmer, and speaking rate. Among these, $F_0$ is most widely used in synthetic speech systems to improve naturalness \cite{F0-2, F0-3, F0-4}, supported by reliable pitch estimation algorithms such as YAAPT \cite{yaapt}, DIO \cite{DIO}, and Harvest \cite{harvest}. 

Despite its importance to human perception, prosody remains underexplored in anti-spoofing. Recent studies \cite{pitchimperfect, ASV_prososdy_AS, AS_SER} have examined prosodic features in isolation \cite{pitchimperfect}, in combination with speaker verification embeddings \cite{ASV_prososdy_AS}, or through emotion recognition features \cite{AS_SER}, achieving performance comparable to or exceeding baselines. These findings underscore the potential of prosody as a valuable yet underutilized cue for spoof detection, especially when explicitly integrated with SSL-based contextual embeddings.

Inspired by this, we aim to make anti-spoofing models \textit{listen like a human} by leveraging prosodic awareness to better distinguish between real and spoofed speech. To this end, we propose \textit{HuLA (Human-Like Listener for Anti-spoofing)}, a two-stage prosody-aware multi-task learning framework. In Stage~1, a pretrained SSL model is fine-tuned on prosody-related tasks (e.g. $F_0$ prediction and voiced/unvoiced (V-UV) classification) using only real speech. This stage is designed to improve the model’s understanding of natural prosodic variation. We extract $F_0$ values at the frame level using a pitch extraction algorithm and use them as reference labels during training. To support the $F_0$ prediction task, we incorporate V-UV frame classification. This helps the model learn voiced/unvoiced positioning within the $F_0$ sequence which is an important aspect of prosodic structure. By learning these patterns in real speech, the model builds a foundation for detecting prosodic deviations in synthetic speech. In Stage~2, the model is jointly optimized for spoof detection and the same prosody tasks on both real and synthetic audio, exposing it to general differences while explicitly leveraging prosodic cues. HuLA consistently outperforms baselines on challenging out-of-domain datasets, including expressive/emotional speech (ASVspoof \cite{ASVspoof2021, ASVspoof2024}, EmoFake \cite{EmoFake}, mixed emotions \cite{mixed_emo}) and cross-lingual data \cite{ADD2022, habla}.

Our key contributions are as follows:
\begin{itemize}
\item We identify imperfect modeling of expressiveness as both a vulnerability of speech synthesis and a discriminative opportunity for anti-spoofing.
\item We highlight the overlooked role of prosody, particularly $F_0$ and voiced/unvoiced patterns, as a discriminative cue for detecting expressive synthetic speech, addressing a key gap in current anti-spoofing research.
\item We introduce HuLA, a two-stage multi-task framework that leverages prosody-awareness for improved spoof detection.
\item We demonstrate strong out-of-domain robustness, showing HuLA’s effectiveness against expressive, emotional, and cross-lingual spoofing attacks.
\end{itemize}
The rest of this paper is organized as follows. Section II reviews related work. Section III examines how expressiveness in synthetic speech presents both a threat and an opportunity for anti-spoofing, motivating our approach. Section IV details the proposed method, while Section V describes the experimental setup. Section VI reports results and explanations, and Section VII provides a discussion. Finally, Section VIII concludes the paper and outlines future directions.

\begin{figure*}[ht!]
    \centering
    \includegraphics[,clip,width=0.8\textwidth]{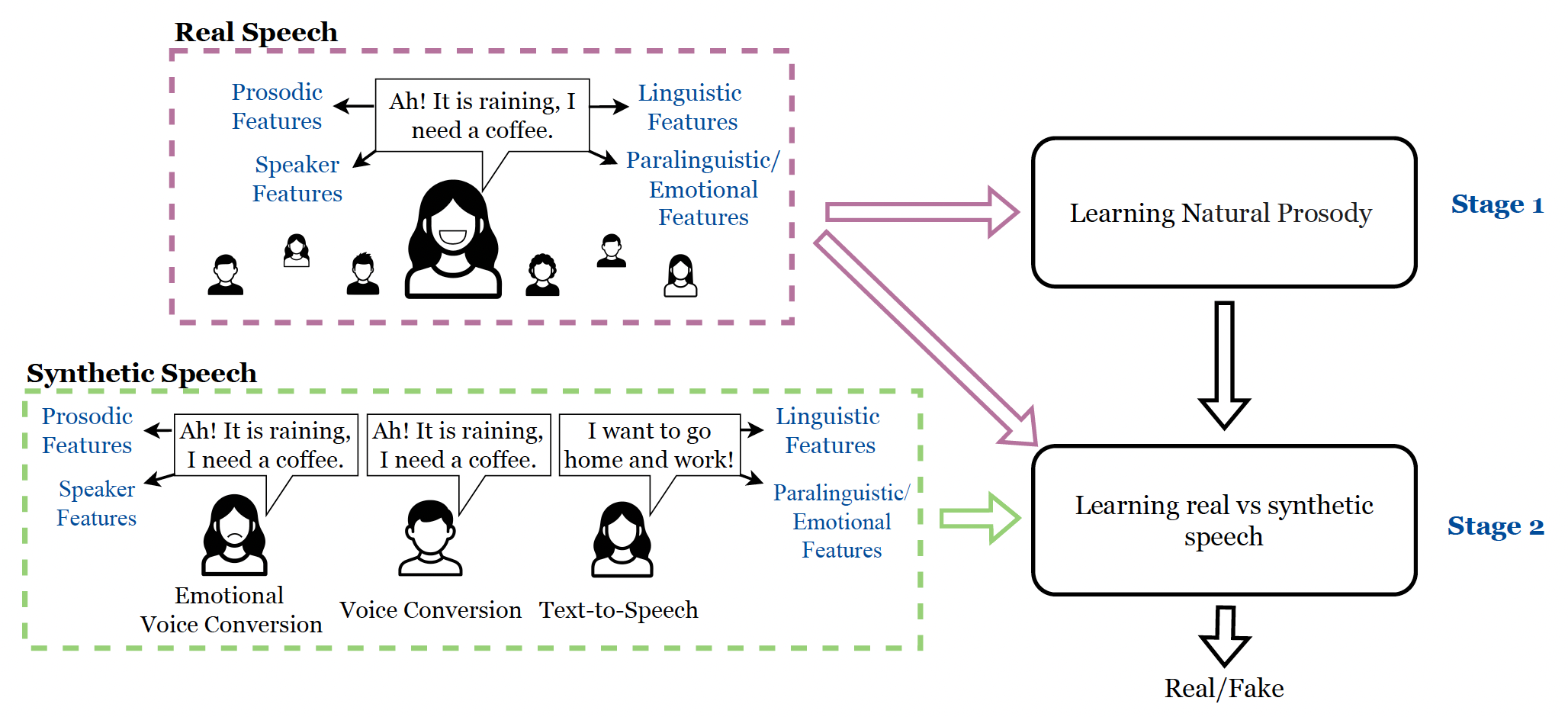}
    \caption{Conceptual overview of our approach. Stage~1 captures the natural prosodic variation of real speech, while Stage~2 leverages this knowledge to distinguish real from synthetic expressive speech.}
    \label{fig:hula_glance}
\end{figure*}

\section{Related Work}
\subsection{Traditional Anti-Spoofing and Datasets}
Anti-spoofing research has been actively driven by community challenges such as ASVspoof and ADD. These initiatives have produced large-scale benchmark datasets including ASVspoof 2019 \cite{ASVspoof2019}, ASVspoof 2021 \cite{ASVspoof2021}, ASVspoof 5 \cite{ASVspoof2024}, ADD 2022 \cite{ADD2022}, and ADD 2023 \cite{ADD2023}. The datasets cover a broad range of attack types such as replay, TTS, VC, and adversarial methods, and emphasize different aspects of spoofing. Some focus on realistic scenarios with channel variability, while others extend coverage to multiple languages. Collectively, these challenges and datasets have provided the foundation for progress in spoof detection research.  

Building on these resources, anti-spoofing methods can be grouped into three major categories: end-to-end models, handcrafted-feature and ssl-based approaches. End-to-end systems such as RawNet2 \cite{RawNet2} and AASIST \cite{AASIST} represent the state of the art, operating directly on raw waveforms and achieving strong benchmark performance. RawNet2 combines a SincNet layer, residual blocks, a GRU, and fully connected layers, while AASIST employs graph attention networks to capture both spectral and temporal artifacts. In contrast, handcrafted approaches extract spectral features such as MFCC \cite{MFCC}, CQCC \cite{CQCC}, Mel-spectrograms \cite{melspec}, or silence proportion \cite{impactofsilenceAS, silenceAS}, followed by supervised classifiers such as LCNN \cite{LCNN}, ASSERT \cite{ASSERT}, or ResNet34 \cite{ResNet34}. However, these models generally struggle with newer datasets that include attacks generated by advanced TTS and VC systems under more realistic conditions. This limitation has motivated a shift toward self-supervised learning (SSL)-based models, which we review in the following subsection.

\subsection{SSL-based anti-spoofing}

Pretrained self-supervised learning (SSL) models have become increasingly popular in anti-spoofing research due to their strong generalization ability and competitive performance. Recent state-of-the-art (SOTA) SSL-based systems \cite{ssl_sls, ssl_rawnet2, ssl-mamba} leverage large-scale pretrained representations to achieve significant improvements over traditional approaches. Commonly used SSL architectures include Wav2Vec 2.0 \cite{wav2vec2}, XLS-R \cite{xlsr}, HuBERT \cite{hubert}, Whisper \cite{whisper}, and WavLM \cite{wavlm}, which are fine-tuned for spoof detection. Several integration strategies have been explored. For example, \cite{ssl_rawnet2} combines XLS-R with a RawNet2 encoder and an AASIST backend, while \cite{ssl_sls} employs XLS-R with a Sensitive Layer Selection (SLS) module, reporting substantial gains on the ASVspoof 2021 DF evaluation track. Both approaches also benefit from data augmentation techniques such as RawBoost \cite{rawboost}. Another recent work \cite{moe_wav2vec2} fuses Wav2Vec 2.0 with a Mixture of Experts (MoE) mechanism and AASIST for classification.  

Despite their success, current SSL-based anti-spoofing models largely overlook prosody, a central cue in human perception of expressiveness and emotion. They do not exploit the imperfect reproduction of prosodic patterns in modern TTS and VC systems, an underutilized discriminative signal with strong potential for spoof detection and the central focus of this work.

\subsection{The Use of Prosody Features in Anti-Spoofing}
Although prosody is imperfectly reproduced in synthetic speech, its use as a feature for spoof detection has been relatively limited. Nonetheless, several studies highlight its potential. For example, ProsoSpeaker \cite{ASV_prososdy_AS} combined prosodic embeddings with speaker verification features and achieved improvements over the baseline. Similarly, \cite{pitchimperfect} showed that a detector based solely on classical prosodic features performed on par with established systems. Another study \cite{prosody_pronunciation_ssl_AS} proposed a cross-dataset framework that fused prosodic and pronunciation features with learned representations, improving generalization.  

Collectively, these works demonstrate that prosody can serve as a valuable cue for spoof detection. However, despite the recent success of self-supervised learning (SSL) models, explicit integration of prosodic modeling with contextual SSL embeddings remains largely unexplored. Addressing this gap forms the core motivation of our work.

\subsection{Summary of Research Gap} %my version is sharper
We identify four major gaps in current spoof detection research:
\begin{itemize}
    \item Existing models are not designed to handle attacks generated with emotional and expressive speech, which limits their robustness in realistic scenarios.  
    \item The imperfect reproduction of expressiveness and emotion in synthetic speech remains an underexploited cue that could be leveraged for detection.  
    \item Prosody, though central to human perception of expressiveness, is underutilized. Most prior work treats it in isolation rather than integrating it with SSL-based embeddings.  
    \item Training datasets are often skewed toward spoofed samples, giving models limited exposure to authentic speech and weakening their ability to learn natural acoustic properties.  
\end{itemize}

This work addresses these gaps by proposing \textit{HuLA}\footnote{Project page: \url{https://aurosweta-jm18.github.io/HuLA/}}, a prosody-aware, two-stage multi-task learning framework that combines SSL representations with explicit prosodic supervision to improve the robustness of anti-spoofing systems.

\section{Expressive Speech Synthesis: A Threat and an Opportunity for Anti-Spoofing}

Recent advances in TTS and VC have enabled increasingly natural synthetic speech with human-like emotion and expressiveness \cite{zkun1,mixed_emo, liu2021reinforcement}. These improvements pose a growing threat to anti-spoofing systems. For example, \cite{emotionAS} investigated emotion-targeted attacks, where adversaries generated high-quality, expressive speech to bypass anti-spoofing models. The results revealed high equal error rates (EERs), with performance varying across both emotions and synthesis models, underscoring the vulnerability of current anti-spoofing systems to emotionally expressive attacks.  

\begin{table}[t]
\centering
\renewcommand{\arraystretch}{1.9}
\caption{Performance of anti-spoofing models trained on ASVspoof 2019 and evaluated on emotional and expressive synthetic speech.}
\begin{tabular}{l|c|c}
\hline
\multirow{2}{*}{\textbf{Model}} & \multicolumn{2}{c}{\textbf{Evaluation (EER\%)}} \\ \cline{2-3}
& EmoFake & ASVspoof 2024 Track 1 \\
\hline
RawNet2~\cite{RawNet2} & 21.71 & 40.67 \\
\hline
AASIST~\cite{AASIST} & 13.64 & 35.53 \\
\hline
\end{tabular}
\label{tab:emoexpon2019}
\end{table}

\begin{table}[t]
\vspace{-2mm}
\centering
\renewcommand{\arraystretch}{1.9}
\caption{Performance of anti-spoofing models fine-tuned on EmoFake and evaluated on emotional and expressive synthetic speech.}
\begin{tabular}{l|c|c}
\hline
\multirow{2}{*}{\textbf{Model}} & \multicolumn{2}{c}{\textbf{Evaluation (EER\%)}} \\ \cline{2-3}
& EmoFake & ASVspoof 2024 Track 1 \\
\hline
RawNet2~\cite{RawNet2} & 11.38 & 44.06 \\
\hline
AASIST~\cite{AASIST} & 4.18 & 36.39 \\
\hline
\end{tabular}
\label{tab:emoexponEF}
\end{table}

To illustrate this challenge, we evaluated state-of-the-art models trained on ASVspoof 2019 \cite{ASVspoof2019}, which does not contain emotional or expressive samples. For evaluation, we used EmoFake \cite{EmoFake}, an emotional anti-spoofing dataset, and ASVspoof 2024 \cite{ASVspoof2024}, the most recent edition of the ASVspoof challenge that includes advanced synthesis techniques capable of generating expressive and realistic speech. As shown in Table~\ref{tab:emoexpon2019}, both RawNet2 and AASIST perform poorly on emotional/expressive attacks. We then fine-tuned these models on the emotional synthetic speech dataset, EmoFake \cite{EmoFake}. Results in Table \ref{tab:emoexponEF} indicate that while performance improves on the EmoFake test partition, performance on ASVspoof 2024 degrades further. These findings suggest that fine-tuning on an emotional corpus (such as EmoFake) improves performance within the same dataset but reduces generalization to other expressive data. This trade-off reflects an architectural limitation: current models often learn dataset-specific artifacts, as also observed in \cite{harderordiff}, and fail to capture more generalizable cues of expressiveness.

These findings raise a deeper question: \textit{Is synthetic expressiveness truly convincing to human listeners?} Studies on human perception of deepfakes suggest otherwise \cite{humanperception-DF}. Human listeners often rely on prosodic cues such as intonation, rhythm, and emphasis to judge authenticity, and such cues remain difficult to reproduce in generated speech. What remains a challenge for machines is often a strength for humans. Current anti-spoofing models, however, do not explicitly incorporate prosody, focusing instead on acoustic features such as log-mel spectrograms or SSL embeddings. This gap presents an opportunity: by explicitly modeling prosody, detectors can learn to identify the same expressive mismatches that humans perceive. Supporting this view, recent work \cite{pitchimperfect} shows that detectors based solely on prosodic features can perform on par with state-of-the-art systems, underscoring the potential of prosody-aware modeling as a complementary path to robustness. 

We believe it is essential for anti-spoofing models to learn the natural prosodic variation of real speech in order to detect synthetic expressive speech produced by advanced synthesis frameworks. Motivated by this, we propose a two-stage training strategy, illustrated in Figure~\ref{fig:hula_glance}. Stage~1 focuses on learning the prosodic patterns of real speech, while Stage~2 leverages this knowledge to capture prosodic deviations and other spoofing cues that distinguish synthetic from genuine speech. In this view, expressive synthetic speech is not only a threat but also an opportunity: its imperfect replication of prosody can serve as a powerful discriminative signal, if models are designed to “listen like a human.”

The proposed anti-spoofing model, \textit{HuLA}, follows this strategy and is described in detail in the next section.

\begin{figure*}[!t]
    \centering
    \subfloat[Multi-task Learning: Stage 1]{
        \includegraphics[width=0.4\textwidth, height=10.5cm]{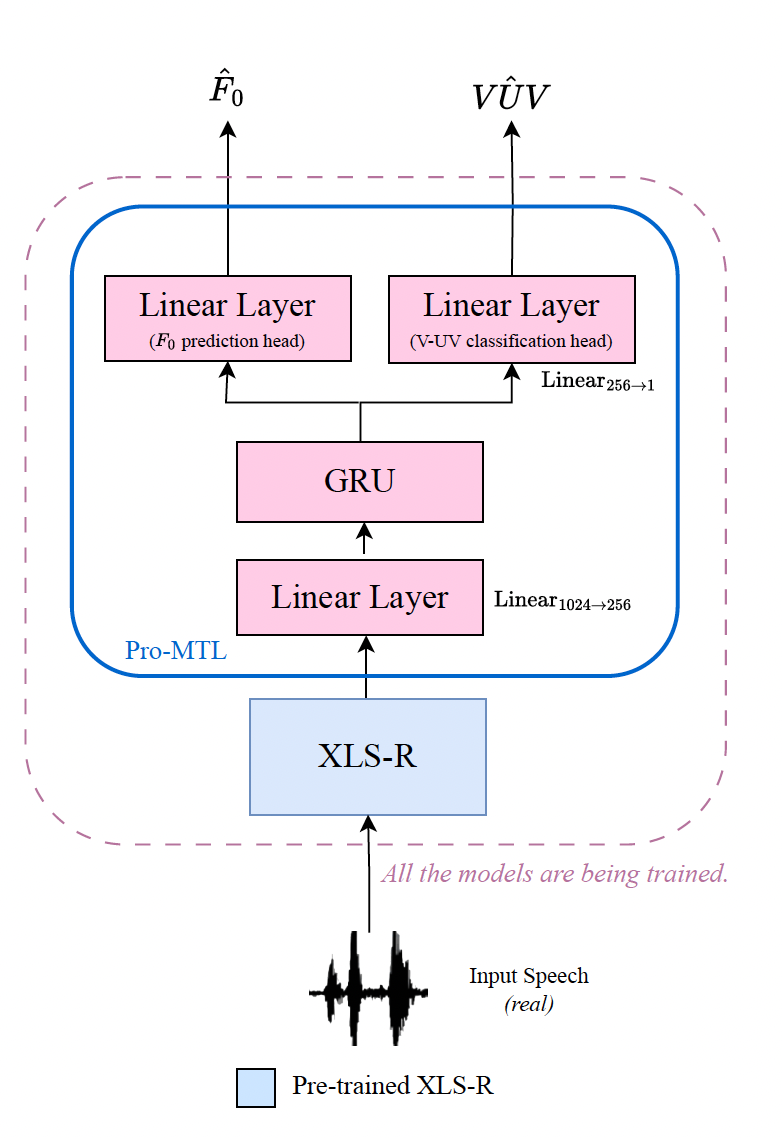}
        \label{fig:subfig1}
    }
    \hfill
    \subfloat[Multi-task Learning: Stage 2]{
        \includegraphics[width=0.55\textwidth, height=10.5cm]{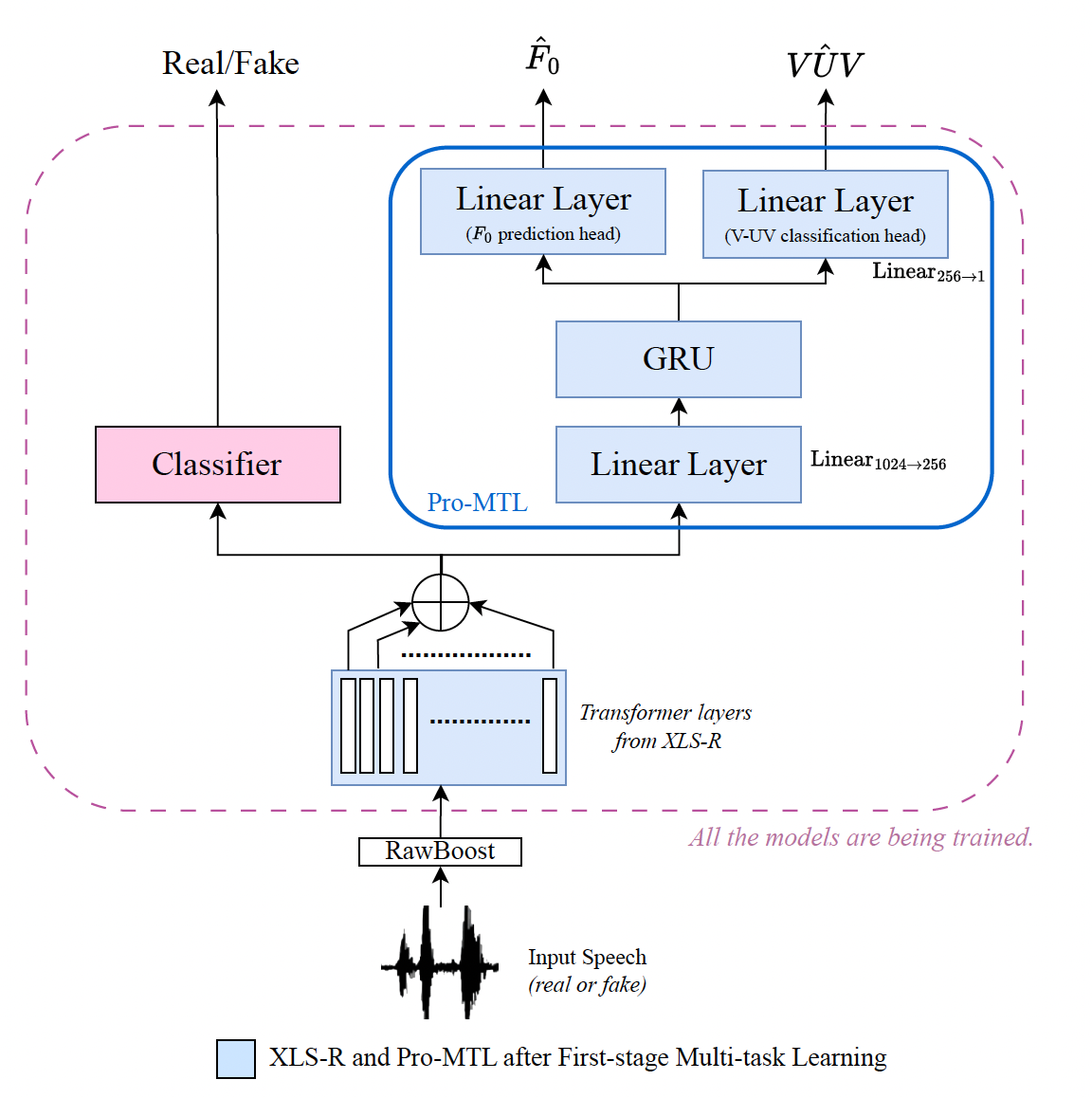}
        \label{fig:subfig2}
    }
    \captionsetup{justification=centering} % This centers the caption text
    \caption{Training phase of HuLA, the proposed prosody-aware multi-task learning method for anti-spoofing. Blue blocks indicate initialization from pretrained models, while pink blocks represents training from scratch.}
    \label{fig:figure1}
    
\end{figure*}

\section{The Proposed Method: HuLA}
Multi-task learning (MTL) \cite{mtl} is a training paradigm in which a model learns multiple tasks simultaneously, often incorporating auxiliary objectives that support the main task. MTL is known to improve generalization by encouraging the model to learn shared representations, and it has been successfully applied in computer vision \cite{mtl_cv} and speech processing \cite{mtl_speech, mtl_speech2, tzeng2025joint}. In anti-spoofing, prior work has shown that MTL can enhance both accuracy and robustness \cite{mtl_as}.

In this work, we introduce \textit{HuLA}, a two-stage MTL framework that makes a self-supervised learning (SSL) backbone explicitly prosody-aware while jointly learning spoof detection. Our design incorporates auxiliary prosody-related tasks—fundamental frequency ($F_0$) prediction and voiced/unvoiced (V-UV) classification—alongside the main classification objective. This enables the model to exploit prosodic variation as a complementary cue to SSL representations, thereby improving the detection of expressive and emotional synthetic speech. Figure~\ref{fig:figure1} provides an overview of the proposed method.

HuLA follows a two-stage training strategy. In Stage~1, the SSL backbone is fine-tuned on real speech using only prosody-related tasks, allowing the model to internalize natural prosodic variation. In Stage~2, the model is trained on both real and spoofed speech, jointly optimizing spoof detection and prosody tasks. This design leverages the prosodic awareness acquired in Stage~1 to better capture mismatches between real and synthetic expressiveness.
\vspace{-3mm}
\subsection{XLS-R Backbone}
We build HuLA on XLS-R \cite{xlsr}, a large-scale SSL model trained on 128 languages that extends the wav2vec~2.0 architecture. A raw waveform $x$ is passed through a convolutional feature encoder to produce latent representations $z \in \mathbb{R}^{T \times 1024}$, where $T$ is the number of frames. A 24-layer transformer network then generates contextualized representations $h$.

The same XLS-R model is used in both training stages. In Stage~1, the output from the final Transformer layer ($H_1 = h^{(24)}$) is used as input to the prosody-related modules. In Stage~2, we compute a weighted sum of all the layers of the Transformer, following \cite{ssl_sls}, to form an aggregate representation $H_2$ that serves as input for both spoof detection and prosody prediction.

We believe that the contextualized representations from XLS-R are rich and informative. Fine-tuning them for prosody-aware auxiliary tasks and spoof detection enhances performance and improves generalization across diverse anti-spoofing datasets.

\vspace{-3mm}

\subsection{Reference Prosody Labels: $F_0$ and V-UV}

As we use $F_0$ prediction and V-UV classification as auxiliary prosody-related tasks on XLS-R, we require frame-level labels for both tasks during training. To obtain these, we use the DIO algorithm \cite{DIO}, a pitch extraction method, and treat its outputs as labels for training.

For each audio sample, the DIO algorithm outputs a sequence of pitch values, denoted as $F_0^{\text{ref}}$:
\begin{equation}
F_0^{\text{ref}} = \left[f_1, f_2, \dots, f_T\right] \in \mathbb{R}^{T \times 1}
\end{equation}

Each value \( f_t \) in the sequence represents the estimated $F_0$ at frame \( t \). The $F_0$ contour of the audio is defined by the full sequence \( F_0^{\text{ref}} \). Voicing is determined based on the $F_0$ values: 
\begin{equation}
v_t = 
\begin{cases}
0 & \text{if } f_t = 0 \ (\text{unvoiced}) \\
1 & \text{if } f_t > 0 \ (\text{voiced})
\end{cases}
\end{equation}

These \( F_0^{\text{ref}} \) and \( VUV^{\text{ref}} \) reference labels are used in both stages of MTL.

\subsection{First-stage Multi-task Learning}

In the first stage of our MTL framework, we aim to guide the model in learning the prosodic characteristics of expressive speech, specifically focusing on $F_0$ prediction and V-UV classification. To achieve this, we fine-tune a pretrained XLS-R model using large-scale real speech data with speaker variability.

\paragraph{Architecture} As illustrated in Figure~\ref{fig:subfig1}, we use the final hidden layer output of XLS-R as input features to the prosody-related task module, referred to as the Pro-MTL module. The Pro-MTL module consists of a linear layer, a GRU, an $F_0$ prediction head, and a V-UV classification head.

First, the input $H_1$ is projected through a linear layer to reduce its dimensionality:
\begin{equation}
L_1 = \text{Linear}_{1024 \rightarrow 256}(H_1)
\end{equation}
The projected features are then passed through a single-layer Gated Recurrent Unit (GRU) \cite{GRU} with a hidden size of 256, which captures temporal dynamics relevant to prosody:
\begin{equation}
G = \text{GRU}_{256}(L_1)
\end{equation}
We choose the GRU due to its ability to efficiently model sequential dependencies. We believe that it's configuration provides a good balance between model capacity and generalization, allowing the network to effectively capture prosody-related patterns.

The GRU output is fed into two parallel heads: one for predicting $F_0$, and another for binary V-UV classification:
\begin{align}
\hat{F}_0 &= \text{Linear}_{256 \rightarrow 1}(G) \\
\hat{VUV} &= \text{Linear}_{256 \rightarrow 1}(G)
\end{align}

\paragraph{Loss Function} We use a weighted multi-task loss that combines:
\begin{itemize}
    \item Mean Squared Error (MSE) loss for $F_0$ prediction,
    \item Binary Cross Entropy (BCE) loss for V-UV classification.
\end{itemize}
\begin{equation}
\mathcal{L}_{\text{total}} = \text{MSE}\left(F_0^{\text{ref}}, \hat{F}_0\right) + \lambda \cdot \text{BCE}\left(VUV^{\text{ref}}, \hat{VUV}\right)
\end{equation}
We empirically set the weight $\lambda = 0.3$.

\paragraph{Normalization} The reference $F_0$ and V-UV labels are generated using the DIO algorithm \cite{DIO}. To improve training stability, we normalize the $F_0$ values using speaker-wise statistics computed on voiced frames only:
\begin{equation}
f_t' = 
\begin{cases}
\displaystyle \frac{f_t - \mu_s}{\sigma_s} & \text{if } f_t > 0 \\
0 & \text{otherwise}
\end{cases}
\end{equation}
Here, $\mu_s$ and $\sigma_s$ denote the speaker-wise mean and standard deviation of the $F_0^{\text{ref}}$ values, computed using only the voiced frames for speaker $s$.

\paragraph{Length Adjustment} During training, we ensure that the model output and target sequences are adjusted by trimming both to the minimum length:
\begin{equation}
T_{\text{adjust}} = \min\left(T_{\text{pred}}, T_{\text{ref}}\right)
\end{equation}
This process prevents frame mismatches during loss computation. By the end of the first-stage MTL, the model develops a strong understanding of prosodic patterns in real speech.

\begin{figure}[t!]
    \centering
    \includegraphics[width=0.5\textwidth, height=5cm]{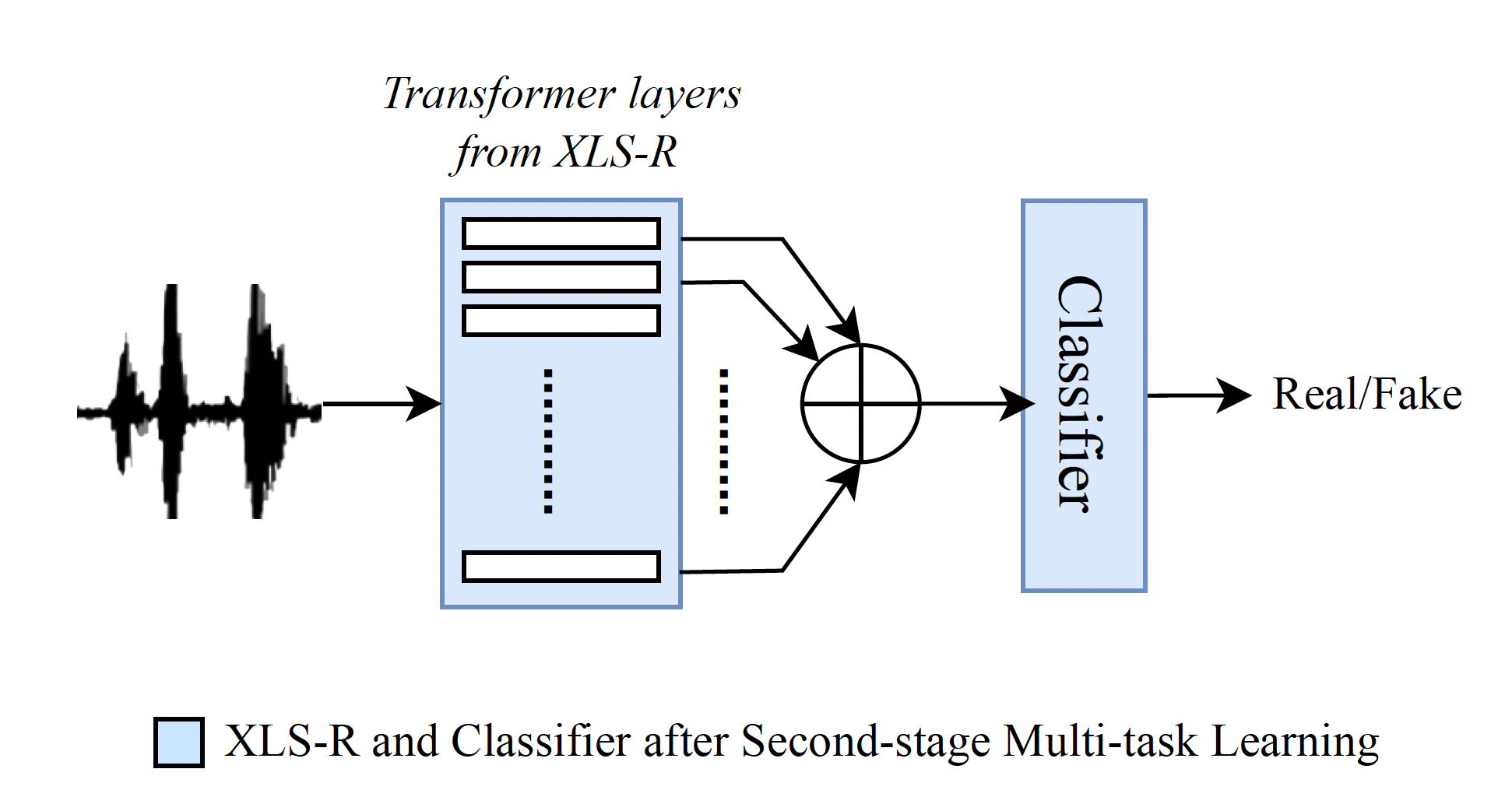}
    \caption{Inference Phase of HuLA}
    \label{fig:figure2}
\end{figure}
\subsection{Second-stage Multi-task Learning}

In the second stage of our MTL framework, we fine-tune the XLS-R model using both real and spoofed speech. Since the model has already been trained to capture prosodic patterns from real speech in the first stage, this phase focuses on learning how prosody differs between real and synthetic speech. In parallel, the model continues to learn other relevant spoof detection cues through updated contextual embeddings obtained after the first-stage MTL. The architecture for the second stage is illustrated in Figure \ref{fig:subfig2}.

We follow the architecture proposed in \cite{ssl_sls}, performing a weighted sum of the outputs from all Transformer layers in XLS-R. This aggregated representation is then passed through the spoof classifier, similar to the setup in \cite{ssl_sls}. In parallel, we retain the Pro-MTL module used in Stage 1 for auxiliary $F_0$ prediction and V-UV classification, as illustrated in Figure~\ref{fig:subfig2}. The same preprocessing pipeline is applied, including speaker-wise normalization of $F_0$ and length adjustment between predicted and reference sequences.

The overall loss is a weighted combination of spoof classification loss and auxiliary prosody task losses:
\begin{align}
\mathcal{L}_{\text{total}} =\; & \mathcal{L}_{\text{CLS}} 
+ \alpha \cdot \Big( \text{MSE}\left(F_0^{\text{ref}}, \hat{F}_0\right) \notag \\
& \quad + \beta \cdot \text{BCE}\left(VUV^{\text{ref}}, \hat{VUV}\right) \Big) 
\end{align}
where $\mathcal{L}_{\text{CLS}}$ is the weighted cross-entropy loss for spoof detection, and $\alpha = 0.4$, $\beta = 0.2$ are empirically chosen weights for the auxiliary tasks.

This joint optimization enables the model to learn from both spoof-relevant and prosody-relevant cues, enhancing robustness under diverse spoofing conditions.

\subsection{Inference Phase}

During the inference phase, we use only the spoof classification component of the model as shown in Figure \ref{fig:figure2}. We load the XLS-R model fine-tuned in the second stage of multi-task learning and retain only the classification head for prediction. In this stage, we discard the auxiliary heads used for prosody-related tasks and retain only the spoof classification head.

\section{Experiments}
\subsection{Dataset} 

We evaluate HuLA across a diverse set of corpora, using two datasets for training and the rest only for evaluation. This design ensures that performance is assessed under realistic, out-of-domain conditions. The datasets are summarized below and detailed statistics are provided in Table~\ref{tab:dataset_stats}.

\begin{itemize}
     \item \textbf{LibriSpeech} \cite{librispeech}: A large corpus of read English speech. We use the \textit{train-clean-100} and \textit{dev-clean} subsets in Stage~1 of multi-task learning to expose the model to diverse speakers and natural prosodic variation in real speech.

    \item \textbf{ASVspoof 2019 (LA)} \cite{ASVspoof2019}: Contains attacks from 19 TTS and VC systems. We use the train and development sets during Stage~2 training and the evaluation partition for performance reporting. This dataset is widely adopted for benchmarking anti-spoofing systems, ensuring fair comparison with prior work.

    \item \textbf{ASVspoof 2021 (LA)} \cite{ASVspoof2021}: An evaluation-only dataset derived from ASVspoof 2019, but transmitted through real telephony systems (VoIP and PSTN) to introduce channel variability. It enables assessment of model robustness under realistic transmission conditions.
    
     \item \textbf{ASVspoof 2024 (ASVspoof 5)} \cite{ASVspoof2024}: The most recent edition of the ASVspoof challenge, incorporating 32 spoofing systems (TTS, VC, and adversarial attacks) and diverse codec/compression conditions. We use the Track~1 evaluation partition to test HuLA against large-scale, state-of-the-art spoofing methods.

      \item \textbf{EmoFake} \cite{EmoFake}: A dataset targeting emotional spoofing attacks. It is constructed using seven emotional VC systems to generate synthetic emotional speech, while bona fide samples are drawn from the Emotional Speech Database (ESD) \cite{ESD}. We use the English-test partition to evaluate HuLA on emotionally expressive fakes.

     \item \textbf{Mixed Emotions} \cite{mixed_emo}: A small-scale dataset designed to study mixed-emotion attacks, where speech conveys blended emotions (e.g., happy+surprise). Synthetic samples are provided by the authors, and bona fide recordings are again taken from the ESD \cite{ESD} corpus.

     \item \textbf{ADD 2022} \cite{ADD2022}: A Mandarin dataset containing both fully and partially fake utterances generated by multiple TTS and VC systems, further degraded with noise and channel effects. We use the Track~1 test set to evaluate cross-lingual spoof detection under realistic conditions.
     
    \item \textbf{HABLA} \cite{habla}: A multilingual evaluation dataset focusing on Latin American Spanish accents (Argentinian, Colombian, Peruvian, Venezuelan, and Chilean). Fake speech is generated using six TTS and VC systems. We use this dataset to assess cross-lingual robustness.
\end{itemize}

\begin{table}[htbp]
\centering
\renewcommand{\arraystretch}{1.9}
\caption{Number of utterances in each dataset.}
\label{tab:dataset_stats}
\begin{tabular}{c|c|c|c}
\hline
\textbf{Dataset} & \textbf{Real} & \textbf{Spoof} & \textbf{Total} \\
\hline
\textit{Train-clean-100 (LibriSpeech)} & 28,539 & -- & 28,539 \\
\textit{Dev-clean (LibriSpeech)}       & 2,703  & -- & 2,703  \\
\hline
\textit{Train (ASVspoof 2019)}         & 2,580  & 22,800  & 25,380 \\
\textit{Dev (ASVspoof 2019)}           & 2,548  & 22,296  & 24,844 \\
\hline
\multicolumn{4}{c}{\textbf{Evaluation}} \\
\hline
ASVspoof 2019 (LA)  & 7,355   & 63,882   & 71,237   \\
ASVspoof 2021 (LA)  & 14,816  & 133,360  & 148,176  \\
ASVspoof 5 (Track 1)  & 138,688 & 542,086  & 680,774  \\
EmoFake  & 3,500   & 14,000   & 17,500   \\
Mixed Emotions  & 160 & 331  & 491  \\
ADD 2022 (Track 1)  & 31,334  & 77,865  & 109,199  \\
HABLA  & 22,816 & 53,000  & 75,816  \\
\hline
\end{tabular}
\end{table}

\subsection{Baseline}
We compare HuLA against several state-of-the-art anti-spoofing models, covering both end-to-end architectures and SSL-based approaches:
\begin{itemize}

     \item \textbf{RawNet2} \cite{RawNet2}: An end-to-end model that operates directly on raw waveforms and has achieved robust performance across ASVspoof benchmarks. We evaluate the official pre-trained model released by the authors, trained on the ASVspoof 2019 dataset, and apply it directly to all evaluation sets in this study. 

    \item \textbf{AASIST} \cite{AASIST}: An anti-spoofing model that employs integrated spectro-temporal graph attention networks to capture both spectral and temporal artifacts in speech. It has consistently ranked among the top-performing systems in recent challenges. For comparison, we use the model trained on ASVspoof 2019, as released by the authors.

    \item \textbf{SSL-SLS} \cite{ssl_sls}: A self-supervised learning–based anti-spoofing system that integrates XLS-R representations with a Selective Layer Selection (SLS) module. The SLS mechanism computes a weighted combination of Transformer layer outputs, emphasizing layers most relevant for spoof detection. Following \cite{ssl_sls}, we replicate their experimental setup: the XLS-R backbone is fine-tuned on the ASVspoof 2019 LA training set, and a validation set is used to monitor convergence and select the best-performing checkpoint.

\end{itemize}

\subsection{HuLA Implementation Details}
For consistency across experiments, all audio samples are either trimmed or zero-padded to approximately four seconds, ensuring uniform input length and frame structure. Frame-level $F_0$ labels are extracted using a frame length of 25\,ms and a frame period of 20\,ms, matching the configuration of XLS-R. We adopt the pretrained XLS-R 300M model \cite{xlsr} as the backbone for all experiments.

Training is carried out in two stages, each for 50 epochs with a batch size of 5 and layer-wise learning rates. In Stage~1, the XLS-R backbone is fine-tuned with a learning rate of $1\times 10^{-6}$, while the Pro-MTL module uses a slightly higher rate of $1\times 10^{-5}$ to allow faster adaptation. No weight decay is applied in this stage to prevent underfitting when learning prosodic patterns from real speech. In Stage~2, the learning rate is kept at $1\times 10^{-6}$ for both XLS-R and the spoof classifier, and $1\times 10^{-5}$ for the Pro-MTL module. A weight decay of $1\times 10^{-4}$ is applied to improve generalization during joint spoof/prosody training. To further enhance robustness, we adopt RawBoost augmentation (Method~3) \cite{rawboost} during Stage~2. This augmentation introduces stationary, signal-independent noise at the waveform level. Prior studies \cite{ssl_sls} have shown that RawBoost improves cross-condition generalization in anti-spoofing, making it a natural choice for our training pipeline.

\begin{table}[t]
\centering
\renewcommand{\arraystretch}{1.9}
\caption{Comparison of baseline and proposed models on the ASVspoof evaluation sets in terms of EER\%.}
\begin{tabular}{l|c|c|c}
\hline
\multirow{2}{*}{\textbf{Model}} & \multicolumn{3}{c}{\textbf{ASVspoof}} \\ \cline{2-4}
 & 2019 LA & 2021 LA & 2024 Track 1 \\
\hline
RawNet2~\cite{RawNet2} & 4.60 & 8.08 & 40.67 \\
\hline
AASIST~\cite{AASIST} & 0.83 & 8.15 & 35.53 \\
\hline
SSL-SLS~\cite{ssl_sls} & 0.56 & 3.04 & 25.43 \\
\hline
HuLA w/o PT & \textbf{0.48} & 1.85 & 23.12 \\
\hline
\textbf{HuLA \textit{(proposed)}} & 0.80 & \textbf{1.38} & \textbf{17.34}  \\
\hline
\end{tabular}
\label{tab:asvspoof}
\end{table}

\begin{table}[t]
\centering
\renewcommand{\arraystretch}{1.9}
\caption{Comparison of baseline and proposed models on the emotional evaluation sets in terms of EER\%.}
\begin{tabular}{l|c|c}
\hline
\multirow{2}{*}{\textbf{Model}} & \multicolumn{2}{c}{\textbf{Emotional datasets}} \\ \cline{2-3}
 & EmoFake & Mixed Emotions \\
\hline
RawNet2~\cite{RawNet2} & 21.71 &  30.00 \\
\hline
AASIST~\cite{AASIST} & 13.64 & 28.12 \\
\hline
SSL-SLS~\cite{ssl_sls} & 8.84 & 16.87 \\
\hline
HuLA w/o PT & 5.24 & 19.37 \\
\hline
\textbf{HuLA \textit{(proposed)}} & \textbf{3.01} & \textbf{16.25} \\
\hline
\end{tabular}
\label{tab:emotions}
\end{table}

\begin{table}[t]
\centering
\renewcommand{\arraystretch}{1.9}
\caption{Comparison of baseline and proposed models on the non-English evaluation sets in terms of EER\%.}
\begin{tabular}{l|c|c}
\hline
\multirow{2}{*}{\textbf{Model}} & \multicolumn{2}{c}{\textbf{Non-English datasets}} \\ \cline{2-3}
  & HABLA & ADD 2022 Track 1 \\
\hline
RawNet2~\cite{RawNet2} & 40.99 & 50.35 \\
\hline
AASIST~\cite{AASIST} & 39.65 & 47.92 \\
\hline
SSL-SLS~\cite{ssl_sls} & 11.58 & 36.93 \\
\hline
\makecell{HuLA\\ w/o PT} & \textbf{8.83} & 33.35 \\
\hline
\makecell{\textbf{HuLA} \\ \textbf{\textit{(proposed)}}}
 & 13.51 & \textbf{32.50} \\
\hline
\end{tabular}
\label{tab:otherLang}
\end{table}

\vspace{-3mm}

\section{Results}
The experiments evaluate the effectiveness of HuLA across multiple benchmarks. We analyze how each stage of our two-stage MTL framework contributes to spoof detection, and demonstrate that explicit prosody modeling improves robustness to expressive, emotional, and cross-lingual attacks.

\subsection{Results on ASVspoof Benchmarks} 
Table~\ref{tab:asvspoof} compares HuLA with strong baselines on three editions of the ASVspoof challenge. End-to-end models such as RawNet2 \cite{RawNet2} and AASIST \cite{AASIST} perform competitively on ASVspoof~2019, but their performance drops sharply on ASVspoof~2021 and ASVspoof~2024. This reflects the increased difficulty of the newer datasets, which introduce channel variability (ASVspoof~2021) and advanced spoofing systems (ASVspoof~2024). The SSL-SLS baseline \cite{ssl_sls} achieves clear gains by leveraging pretrained representations, yet still struggles with ASVspoof~2024, where expressive and natural synthetic speech is included.

HuLA consistently outperforms all baselines. The one-stage version (without Stage~1 pretraining) achieves the best results on ASVspoof~2019 (0.48\% EER), while the full two-stage framework generalizes better to newer conditions, reaching 1.38\% EER on ASVspoof~2021 and 17.34\% EER on ASVspoof~2024. These findings demonstrate that prosody-aware supervision complements SSL embeddings, enabling HuLA, although trained only on ASVspoof 2019, to generalize effectively to advanced synthesis techniques (including expressive samples) as well as realistic channel variability, outperforming prior systems.
\vspace{-3mm}
\subsection{Results on Expressive Synthetic Speech}
ASVspoof~2024 includes a number of modern TTS and VC systems known for producing expressive, human-like speech (e.g., \cite{expressive_VITS, expressive_xtts, expressive_fastpitch, expressive_glowtts, expressive_gradtts, expressive_starganv2, expressive_diffVC, expressive_VAE_GAN}). As shown in Table~\ref{tab:asvspoof}, such attacks remain challenging for existing state-of-the-art baselines: RawNet2, AASIST, and SSL-SLS all experience notable degradation. 

HuLA achieves substantially better results (17.34\% EER), demonstrating that explicitly modeling prosody enables the system to detect subtle mismatches between synthetic and natural expressiveness that other models overlook. In addition, the two-stage variant surpasses the one-stage version, confirming that learning prosodic variation from real speech in Stage~1 provides a stronger foundation than training only in mixed real and spoofed speech in Stage~2.

\subsection{Results on Emotional Synthetic Speech}
Emotional synthetic speech poses a particularly serious challenge for anti-spoofing \cite{emotionAS}. As shown in Table~\ref{tab:emotions}, baseline models such as RawNet2 \cite{RawNet2} and AASIST \cite{AASIST} perform poorly on the EmoFake dataset, which contains emotionally expressive synthetic speech. The SSL-based baseline achieves stronger results, but HuLA without pretraining further improves performance. The full two-stage HuLA model provides the best results overall, indicating that each stage of the framework contributes to increased prosody-awareness. These findings suggest that HuLA is especially effective for prosody-rich attacks such as emotional speech.

While EmoFake focuses on single-emotion attacks, real-world speech often blends multiple emotions \cite{mixed_emo}. To examine this case, we also evaluate on a mixed-emotion dataset containing combinations such as \textit{surprise+happy} and \textit{happy+angry}. As shown in Table~\ref{tab:emotions}, RawNet2 and AASIST are highly vulnerable to such attacks, while SSL-SLS shows notable improvement. HuLA without pretraining underperforms SSL-SLS in this case, but the full two-stage HuLA achieves the best results. This experiment underscores the value of learning prosodic variation from real speech in Stage~1, which equips HuLA to handle complex emotional mixtures where multiple prosodic patterns interact.
\vspace{-3mm}
\subsection{Results on Cross-Lingual Synthetic Speech}

Cross-lingual evaluation investigates whether models trained in one language can generalize to others. This is particularly challenging since the model is tested on a language it has never encountered during training. As an additional evaluation, we assess HuLA in this setting, using spoofing datasets in languages other than English while training all models exclusively on English data. Prior work has shown that anti-spoofing models often achieve limited cross-lingual generalization \cite{cross-lingual_as, cross-lingual_as_2, habla}. Building on these insights, we test whether HuLA’s prosody-aware design offers an advantage in this more difficult scenario.

Table~\ref{tab:otherLang} reports results on HABLA, a Latin American Spanish dataset. HuLA achieves the best performance, outperforming both end-to-end and SSL-based baselines. 
Although both HuLA and SSL-SLS rely on the multilingual XLS-R backbone, HuLA’s superior results highlight the added value of explicit prosody modeling. Interestingly, HuLA without Stage~1 pretraining outperforms the full model, likely because pretraining on English-only speech limits generalization for Spanish.

We also evaluate on ADD 2022 Track~1, a Mandarin dataset with fully fake utterances. Here, RawNet2 and AASIST perform poorly, while SSL-SLS achieves stronger results. Both HuLA variants surpass SSL-SLS, with the full two-stage HuLA achieving the best performance. While HuLA was not specifically designed for cross-lingual attacks, its prosody-aware design proves beneficial in this setting. Performance remains strong on Spanish but is more limited on Mandarin, reflecting the difficulty of transfer to typologically distant languages. However, HuLA outperforms the baselines in both cases. These findings suggest that prosody-aware modeling is a promising path toward cross-lingual robustness and motivate further investigation in this direction.

\section{Discussion}

Our experiments demonstrate the effectiveness of HuLA, a two-stage MTL framework that improves spoof detection through explicit prosody modeling. Although trained only on ASVspoof 2019, which lacks the diversity and realism of recent attacks, HuLA generalizes well across datasets that differ substantially from the training domain. Several of these sets include expressive and emotional synthetic speech, which typically fool state-of-the-art baselines. Prior work \cite{harderordiff} shows that anti-spoofing models often overfit to dataset-specific artifacts, limiting cross-domain robustness. In contrast, HuLA benefits from prosody-aware training, which equips the model to detect mismatches in expressiveness that are not dataset-dependent. This aligns with our design principle of 
\textit{listening like a human}: just as listeners use prosodic cues to judge naturalness, HuLA leverages prosodic variation in both real and spoofed speech to capture subtle differences in expressiveness. %\textcolor{blue}{Section 3 further showed that fine-tuning baselines on emotional corpora does not yield strong generalization, revealing an architectural limitation of current models. HuLA overcomes this limitation, generalizing well across emotional, expressive, and cross-lingual datasets. Notably, it outperforms baselines even when they are fine-tuned on emotional data, despite being trained only on ASVspoof 2019.}

Ablation results suggest that HuLA without pretraining (denoted as HuLA w/o PT) already provides substantial improvements over strong baselines, underscoring the value of prosody-aware supervision. Stage~1 pretraining on 100 hours of real speech then delivers further gains, leading to the best overall performance. These findings suggest that pretraining on large, diverse corpora of real speech can enhance detection ability. Performance on emotion attacks further demonstrates the importance of modeling prosodic variation: HuLA effectively handles challenging cases where different emotional states introduces additional variability in acoustic patterns.

Finally, our cross-lingual evaluations reveal both opportunities and limitations. Although trained only on English, HuLA achieves strong performance on Spanish (HABLA) but shows more limited effectiveness on Mandarin (ADD 2022). In both cases, however, HuLA outperforms the baselines. These consistent gains across languages highlight the potential of prosody-aware anti-spoofing for multilingual settings. In future work, we will explore methods that exploit core prosodic patterns to achieve stronger generalization across diverse languages.

\section{Conclusion}
We presented HuLA, a prosody-aware, two-stage multi-task learning framework for spoof detection. Our approach is motivated by the observation that emotions and expressiveness, while challenging for anti-spoofing, also provide discriminative cues that humans instinctively rely on. By explicitly modeling prosodic features such as $F_0$ and voiced/unvoiced classification alongside SSL embeddings, HuLA learns to “listen like a human.” In Stage~1, the model acquires prosodic variation from real speech, and in Stage~2 it jointly optimizes spoof detection and prosody tasks using both real and synthetic speech. Extensive experiments demonstrate that HuLA outperforms strong baselines on diverse datasets, including recent ASVspoof editions, emotional and expressive speech, and cross-lingual corpora. These results show that imperfect replication of prosody in synthetic speech can be exploited as a powerful signal for detection. Future work will extend HuLA by incorporating richer prosodic and paralinguistic features, exploring emotion-aware modeling across diverse affective states, and investigating strategies for stronger multilingual generalization.

\section{Acknowledgment}
This work was supported by the National Science Foundation (NSF) CAREER Award IIS-2533652. The authors also thank the Johns Hopkins University Data Science and AI Institute (DSAI) for support through a faculty startup package.  

\bibliographystyle{IEEEtran}
\bibliography{mybib}

\begin{IEEEbiography}
[{\includegraphics[width=1in,height=1.25in,clip,keepaspectratio]{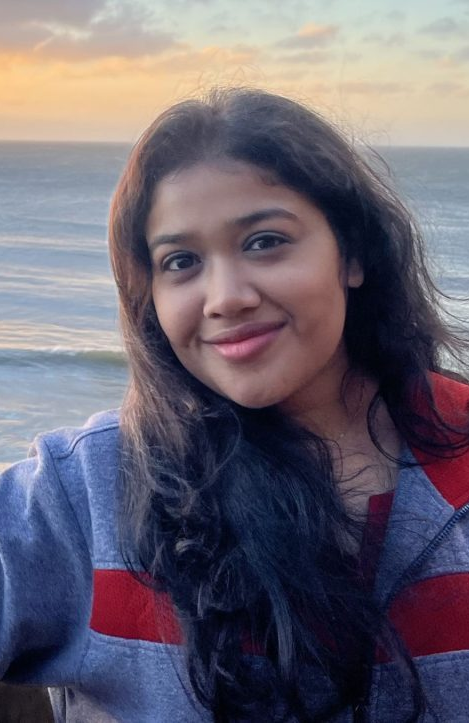}}]{Aurosweta Mahapatra} received the B.Tech. degree in Electronics and Telecommunication Engineering from Kalinga Institute of Industrial Technology, Odisha, India, in 2022, and the M.S. degree in Electrical and Computer Engineering from the University of California, Los Angeles (UCLA), in 2024. She is currently pursuing the Ph.D. degree in Electrical and Computer Engineering at Johns Hopkins University, where she is affiliated with the Center for Language and Speech Processing (CLSP). Her research focuses on secure speech technologies, with an emphasis on developing robust anti-spoofing systems.
\end{IEEEbiography}

\begin{IEEEbiography}
[{\includegraphics[width=1in,height=1.25in,clip,keepaspectratio]{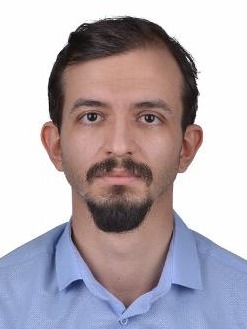}}]{Ismail Rasim Ulgen} (Student Member, IEEE) received the B.S. and M.S. degrees in Electrical and Electronics Engineering from Bogazici University, Istanbul, Turkey, in 2022. He is currently pursuing the Ph.D. degree in Electrical and Computer Engineering at Johns Hopkins University, where he is affiliated with the Center for Language and Speech Processing (CLSP). His research interests include speech synthesis, evaluation, speaker verification and emotion. 
\end{IEEEbiography}

\begin{IEEEbiography}[{\includegraphics[width=1.1in,height=1in,clip,keepaspectratio]{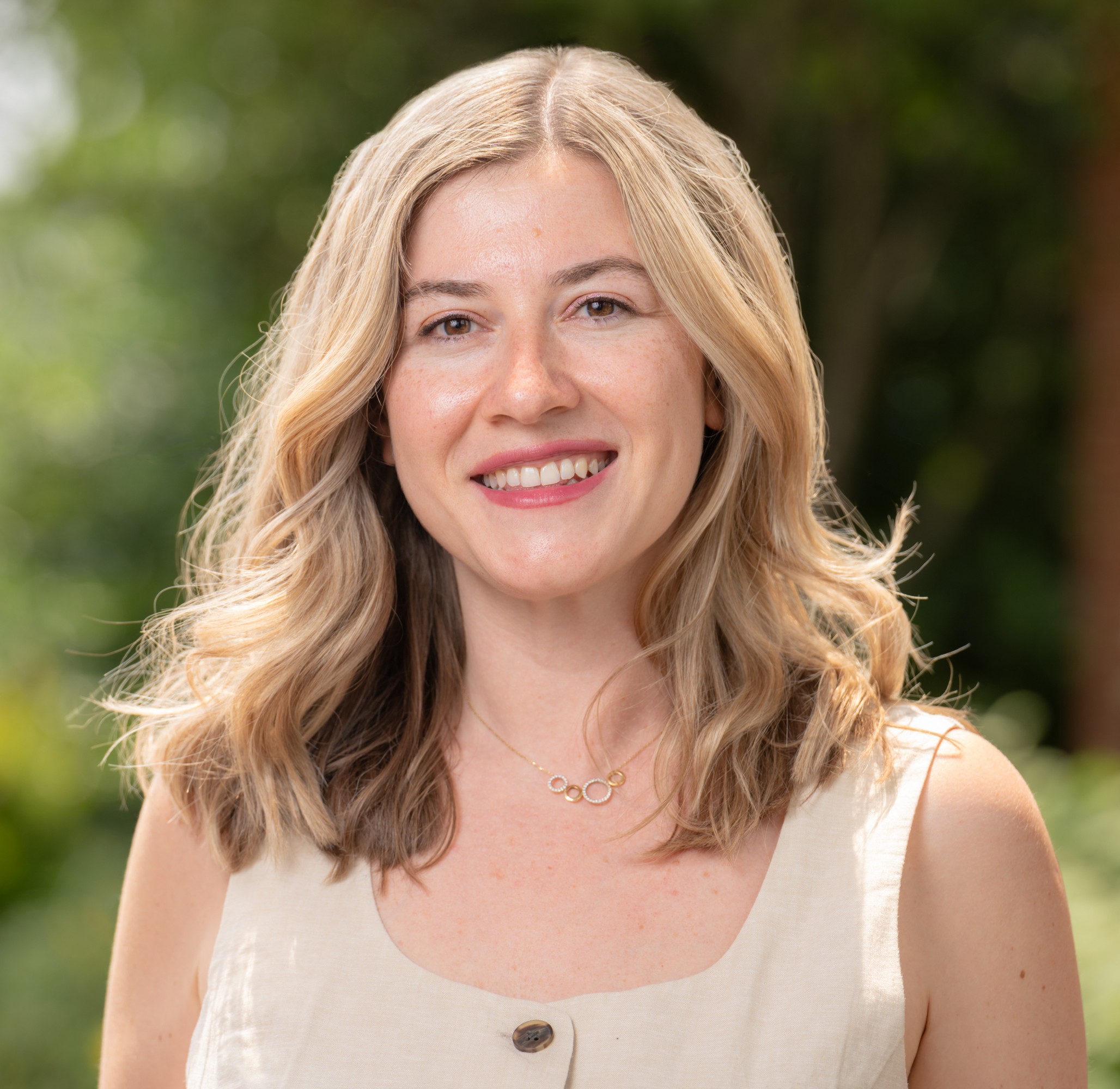}}]{Berrak Sisman} (Senior Member, IEEE) is an Assistant Professor in the Department of Electrical and Computer Engineering at Johns Hopkins University, where she is affiliated with the AI-X Bloomberg Distinguished Professorship Cluster, the Center for Language and Speech Processing (CLSP), and the Data Science and AI Institute (DSAI). She received the Ph.D. degree in Electrical and Computer Engineering from the National University of Singapore, with visiting research appointments at the University of Edinburgh, U.K., and the Nara Institute of Science and Technology (NAIST), Japan. From 2022 to 2024, she was a tenure-track faculty member at the University of Texas at Dallas. Dr. Sisman is a recipient of the NSF CAREER Award (2024), the Amazon Faculty Award (2022), and the Singapore Ministry of Education Tier~2 Grant (2021). She also received the A*STAR Singapore International Graduate Award (2016–2020). She is an elected member of the IEEE Speech and Language Processing Technical Committee and serves as an Associate Editor for the IEEE Transactions on Affective Computing. She is the Technical Program Co-chair for Interspeech 2026 and General Co-Chair for Interspeech 2028. She was elected to the ISCA Board for the 2025–2029 term. Her research interests include speech processing, speech synthesis, voice conversion, deepfake detection and affective computing.
\end{IEEEbiography}
\vfill

\end{document}